\documentclass{article}[15pt]

\usepackage{amsmath,epsfig,color,calc,rotating}
\usepackage{graphics}
\usepackage{wrapfig}
\usepackage{ulem}

\usepackage{natbib}
\bibpunct{(}{)}{,}{a}{}{;}

\begin{document}

\title{
Using Volcano Plots and Regularized-Chi Statistics
in Genetic Association Studies
\author{
Wentian Li$^1$, Jan Freudenberg$^1$, Young Ju Suh$^2$, Yaning Yang$^3$ \\
{\small \sl 1. The Robert S. Boas Center for Genomics and Human Genetics,  The Feinstein Institute }\\
{\small \sl for Medical Research, North Shore LIJ Health System, Manhasset, 350 Community Drive, NY 11030, USA.}\\
{\small \sl 2. Department of Biostatistics ,
School of Medicine, Inha University, Incheon, KOREA} \\
{\small \sl 3. Department of Statistics and Finance,
University of Science and Technology of China, Anhui 230026, Hefei, CHINA} \\
}
\date{}
}
\maketitle  
\markboth{\sl Li, Freudenberg, Suh, Yang}{\sl Li, Freudenberg, Suh, Yang }

\large

\begin{center}
{\bf Abstract}
\end{center}

\normalsize

Labor intensive experiments are typically required to identify the causal 
disease variants from a list of disease associated variants in the genome. 
For designing such experiments, candidate variants are ranked by
their strength of genetic association with the disease. However, the two commonly used 
measures of genetic association, the odds-ratio (OR) and $p$-value, may rank 
variants in different order. To integrate these two measures into a single analysis, 
here we transfer the volcano plot methodology from gene expression analysis to
genetic association studies. 
In its original setting,
volcano plots are scatter plots of fold-change
and $t$-test statistic (or $-$log of the $p$-value), with the latter being more 
sensitive to sample size.  In genetic association studies, 
the OR and Pearson's chi-square statistic (or equivalently its square 
root, chi; or the standardized log(OR)) can be  analogously used in a
volcano plot, allowing  for their visual inspection. 
Moreover, the geometric interpretation of 
these plots leads to an intuitive method for filtering results by a 
combination of both OR and chi-square statistic, which we term 
``regularized-chi". This method selects associated markers by a smooth 
curve in the volcano plot instead of the right-angled lines which
corresponds to independent cutoffs for OR and chi-square statistic. 
The regularized-chi incorporates relatively more signals from 
variants with lower minor-allele-frequencies than chi-square test statistic. 
As rare variants tend to have stronger functional 
effects, regularized-chi is better suited to the task of prioritization of 
candidate genes.

\vspace{0.3in}


\newpage

\large
\section*{Introduction}

\indent

Volcano plots are graphical tools that are commonly used in the
analysis of mRNA expression levels as obtained from microarray
technology \citep{gibson,cui1,wli-review}. In principle, volcano plots 
are scatter plots, with each point representing a probe set or a gene,
and the $x$-coordinate being the (log) fold-change (FC) and $y$ 
being the $t$-statistic or $-\log_{10}$ of the $p$-value from 
a $t$-test. The reason for the popularity of volcano plot in 
microarray data analysis is due to its simultaneous display of 
two, albeit correlated, pieces of information -- fold-change 
and $t$-statistic.
Ranking genes by fold-change and by $t$-test does not necessarily
lead to the same order in the differential expressed gene list, 
and can give rise to different biological conclusions.

However, there is a fundamental relationship between log-fold-change 
and $t$-statistic: while $\log$(FC) is a measure of the magnitude 
of a ``signal", the $t$-statistic is approximately $\log$(FC)
divided by its standard error, i.e., a signal-to-noise ratio 
\citep{zhang,wli-review}. This means that the log(FC) is an unstandardized 
measure of differential expression, whereas $t$-statistic is 
a noise-level-adjusted standardized measure. 
The distinction between the two types of measures of differential
expression has parallels to the long standing discussions in behavioral science, 
psychology, epidemiology, meta-analysis, and engineering under
the theme of ``effect size" \citep{cohen}. As one possible strategy to
address this issue, volcano plots display both measures simultaneously.

In genetic association study, there has been a similar issue
on deciding which measure of association is more useful:
odds-ratio (OR) or $\chi^2$ test (either the chi-square
test statistic or the $p$-value from $\chi^2$ test) \citep{wli-bib}. 
Currently, most association analyses apply a $\chi^2$ test
as the primary single-nucleotide-polymorphism (SNP)
selection criterion in the initial screening, 
and use the OR as a secondary measure in a re-examination.
However, the distinction between the role of two measures
and their connection has not always been explained. Because
selecting candidate SNPs and regions from the first stage
for the replication stage is of great practical importance, 
one would like to add more information in the screening stage.
We believe the application of volcano plots can be beneficial
towards this goal.

One particular advantage of volcano plot in microarray
analysis is that it provides a natural context in
addressing ``joint gene filtering" \citep{zhang,wli-review},
which are the measures of differential expressions using
both log-FC and $t$-statistic. In comparison, the
{\sl ad hoc} selection criterion such as ``FC $> 1.5$, {\sl and}
$p$-value $< 10^{-3}$" could be called ``double gene filtering" 
\citep{zhang,wli-review}. In a volcano plot, the discriminant 
lines for double filtering are right-angled lines formed
by the vertical and the horizontal lines, whereas those 
in joint filtering are smooth curves. The
well known significance analysis of microarray (SAM)
\citep{sam,sam-docu} has a discriminant line in the form
of $y= c_0 + c_1/tan(\theta)$, where $\theta$ is the angle
between the $y$-axis and the line connecting the gene point and the plot origin, 
\citep{wli-review}.

Our goal in transferring volcano plots from expression analysis
to genetic association analyses is to find SNP-filtering criteria that
incorporate information from both OR and $\chi^2$ test
results.  This effort may help the prioritization of SNPs
and chromosome regions in a genome-wide association study \citep{cantor}. 
Prioritization of candidate genes has wide application in every ``omics" 
field \citep{tranch,moreau}.  Usually, the prioritization strategies rely 
on external information of gene products, such as protein-protein 
interaction \citep{pattin} and pathways \citep{chen,kwang,peng}.
In contrast, our approach is purely statistical, with the underlying assumption
that OR may provide more biological information than $\chi^2$-test $p$-value
in a realistic setting (finite sample size and sample heterogeneity).  

The questions that are discussed in this paper are as follows:
(i) How does the minor allele frequency of markers appear in 
a volcano plot? (ii) How can one choose the penalty term
in a regularized-chi statistic for genetic data, and is the choice of 
this term important? 
(iii) Do any other unstandardized and standardized measurements
exist that may be used for $x$- and $y$-axes in the volcano plot, 
besides OR and $\chi^2$ test statistic? 

\section*{Methods and Materials}
\indent

{\bf Unstandardized and sample-size-insensitive measures of differential
allele frequencies:} Denote the $2 \times 2$ 
allele count contingency table as $\{ n_{ij} \}$ ($i,j$= 1,2) 
where row $i$ is the case ($i=1$) or control ($i=2$) label,
and column $j$ is the minor ($j=1$) or major ($j=2$) allele label.
The minor-allele-frequency (MAF) in the case (control) group
is $p_1= n_{11}/(n_{11}+n_{12})= n_{11}/n_{1*}$
($p_2=n_{21}/n_{2*}$). The major-allele-frequencies are
$q_1=1-p_1$ ($q_2= 1-p_2$) for the case (control) group.

One can define several ``differential allele frequency"
measures that are insensitive to the sample size, for instance,
the odds-ratio (OR)  which is defined as 
$n_{11}n_{22}/(n_{12} n_{21})$ or $ p_1q_2/(p_2q_1)$. 
It is well known that the log-transformed OR,
$\log(OR)= \log (n_{11} n_{22}) - \log (n_{12} n_{21}) $, approximately follows
a normal distribution \citep{woolf}.
Another unstandardized measure is the
direct calculation of allele frequency difference between
the case and control group: 
$d_{MAF}= p_1-p_2= n_{11}/n_{1*}- n_{21}/n_{2*}
= (n_{11}n_{22}- n_{12}n_{21})/(n_{1*}n_{2*})$.
Furthermore, Wright's fixation index, 
$F_{st}= 1- (p_1q_1+p_2q_2)/(2\overline{p} \cdot \overline{q}) 
=1- 0.5n^2( n_{11}n_{12}/n_{1*} + n_{21}n_{22}/n_{2*})/(n_{*1}n_{*2})$
($\overline{p}=(p_1+p_2)/2, \overline{q}=(q_1+q_2)/2=1-\overline{p}$)
provides an unstandardized measure of differential allele
frequency\citep{wright}. $F_{st}$ is a measure of allele frequency
difference between two subpopulations that is used in population
genetics and estimates the  proportion of variations explained by
population stratification. Here we assume for simplicity the two subpopulations are case 
and control population, with a 50:50 mixing ratio.

{\bf Standardized and sample-size-sensitive measures of differential
allele frequencies:} 
As a standardized and sample-size sensitive measure of 
differential allele frequency, the $\chi^2$ statistic
($n=n_{**}= \sum_{ij} n_{ij}$) (e.g., \citep{yates,wli-apbc07})
\begin{equation}
X^2= \frac{(n_{11}n_{22}-n_{12}n_{21})^2 n}{ n_{1*}n_{2*}n_{*1}n_{*2}}
= n (p_1-p_2)^2 \frac{n_{1*}n_{2*}}{(p_1n_{1*}+p_2n_{2*})(q_1n_{1*}+q_2n_{2*})}
\end{equation}
is clearly proportional to sample size $n$, in the asymptotic limit,
given $p_1 \ne p_2$. Alternatively,
log-OR itself can be standardized by its standard error, 
$\log(OR)/SE(\log(OR))$, where \citep{woolf}
\begin{equation}
\label{eq-woolf}
SE(\log(OR)) = \sqrt{ \frac{1}{n_{11}} + \frac{1}{n_{12}} + 
 \frac{1}{n_{21}} + \frac{1}{n_{22}} }
=\sqrt{\frac{1}{p_1q_1n_{1*}}+ \frac{1}{p_2q_2n_{2*}}}.
\end{equation}
The standardized $\log(OR)$ is of the form of a Wald statistic.

{\bf A SNP dataset for illustration:} For illustration purpose, 
we first use a genotyping 
data of an autoimmune disease on one chromosome only (chromosome 6) 
with 809 cases and 505 controls \citep{kRA}. 
The initial 38735 SNPs are reduced 
to 35855 SNPs by requiring that both the case and the control 
group to have at least one copy of the minor allele
(which allele is the minor is defined by the control group). This 
removes many SNPs with minor allele frequency (MAF) less than 0.05,
though many low-MAF SNPs remain. These data are
used in Fig.\ref{FIG1}, Fig.\ref{FIG2}, and Fig.\ref{FIG4}.

{\bf Genome-wide association study (GWAS) case-control data
for type 2 diabetes:} We use The Wellcome Trust Case Control
Consortium (WTCCC) data for the type 2 diabetes (T2D), with 1924 
T2D cases and 2938 controls \citep{wellcome,t2d}. Differing from 
several other autoimmune diseases including rheumatoid arthritis
and type 1 diabetes, there is no major susceptibility locus in
MHC region for type 2 diabetes.
The genotyping was carried out by the Affymetrix GeneChip 500k array.
There are 459,446 autosomal SNPs which had already passed a 
quality control (QC) procedure by WTCCC.  

We impose a further filtering criterion: (i) $p$-value for testing 
unbiased typing ratio is larger than $10^{-4}$; (ii) $p$-value for testing 
Hardy-Weinberg equilibrium in both the control and the case group 
is larger than $10^{-4}$; (ii) MAF in both control and case 
group is larger than 0.005.  This reduces the number of SNPs to 388,023
(84.45\% of the original number). 

Our MAF criterion is more relaxed than the 0.01 used
in early analysis of these data \citep{wellcome,t2d},
which leads to the inclusion of more rare variants. 
Although violation of HWE in the case group might be 
considered as part of disease signal 
\citep{feder,weir,song,wli-apbc07,wli-cbc,zheng}, we noticed
that in this data, it actually leads to SNPs with genotype
distribution inconsistent between the case and the control group
(e.g. the SNP rs3777582 on chromosome 6). This data is used in Fig.\ref{FIG3}.

\section*{Results}
\indent

{\bf MAF of SNPs and angle in the volcano plot:} 
To evaluate the role of the MAF of SNPs, we are using the volcano plots with 
$x=\log(OR) = \log (n_{11} n_{22}) - \log (n_{12} n_{21})$,
and $y=\log(OR)/SE(\log(OR))$.
In microarray analysis, the angle $\theta$ between the $y$-axis
and the line linking a gene dot and the origin is
directly related to the standard error of the log-fold-change\citep{wli-review},
$\tan(\theta)=SE(\log(FC))$, and SE in turn is roughly the
standard deviation of log-expression level divided
by $\sqrt{n}$. Consequently, points closer to the $y$-axis
are genes with low-variances. There is a parallel situation here
by replacing fold-change with odds-ratio.

Using the formula in Eq.(\ref{eq-woolf}), we can write SE(log($OR$)) as
(assuming equal number of case and control samples: $n_{1*}=n_{2*}=n/2$):
\begin{equation}
\label{eq-angle}
SE(\log(OR)) = \sqrt{\frac{2}{n}}
\sqrt{\frac{2+(1/p_2+1/q_1)d_{MAF}}
{p_2q_1(1+d_{MAF}/p_2)(1+d_{MAF}/q_1)}}
\end{equation}
If MAF is approaching zero (e.g. $p_2 \rightarrow 0$),
then $SE(\log(OR)) \sim 1/\sqrt{p_2} \rightarrow \infty$ 
and $\theta \rightarrow 90^o$.
These are the points close to the $x$-axis which can have
any OR values.

\begin{figure}[t]
\begin{center}
  \begin{turn}{-90}
  \epsfig{file=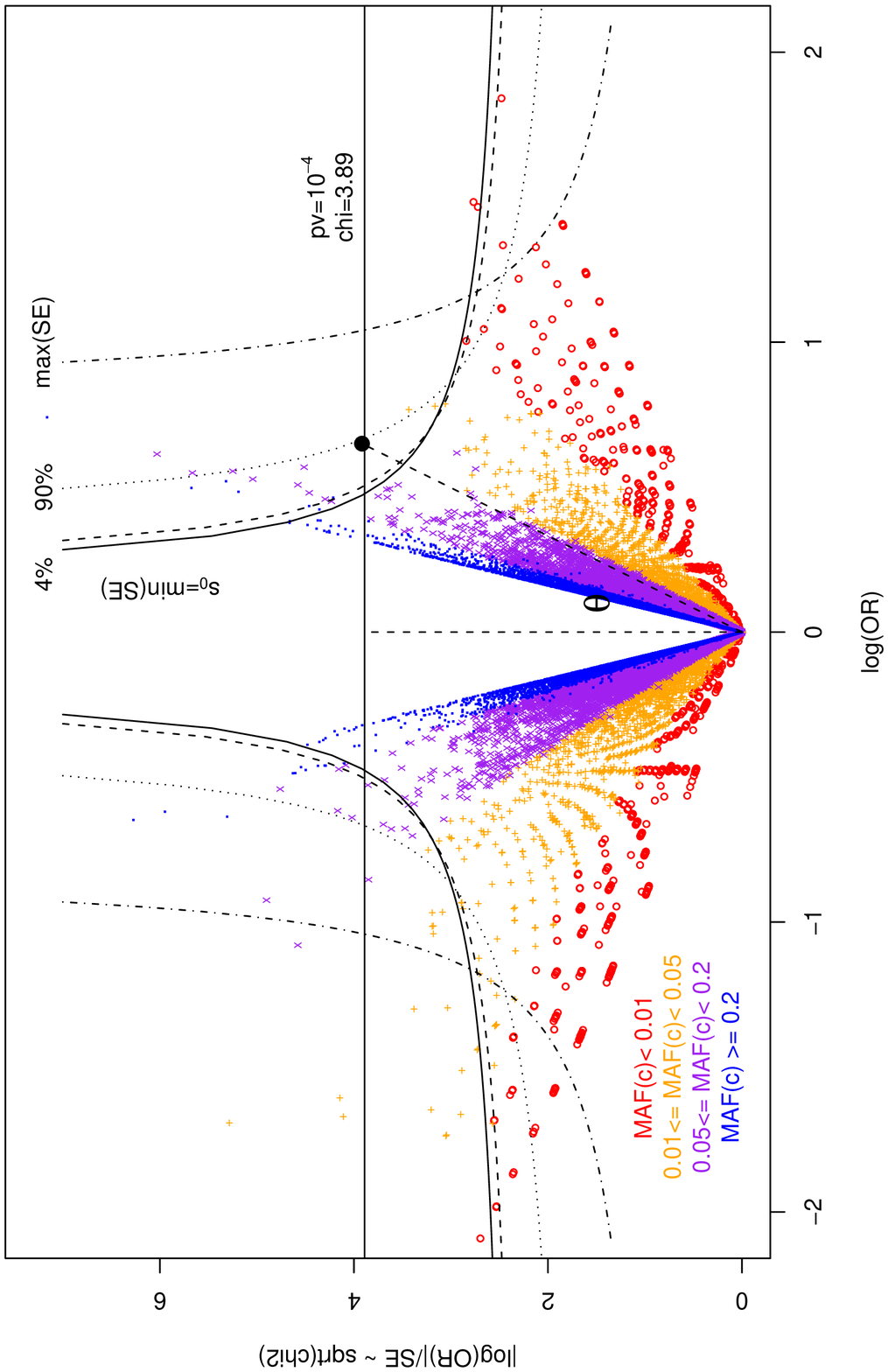, width=9cm}
  \end{turn}
\end{center}
\caption{
\label{FIG1}
}
\end{figure}

Fig.\ref{FIG1} shows how the angle $\theta$ stratifies
SNPs with different MAFs. The SNPs colored with red, orange, purple, blue
correspond to those with control MAF $<0.01$, (0.01, 0.05), (0.05, 0.2),
and $>0.2$. Note that the quality control step has already removed
SNPs with very low MAFs and the control MAFs are greater than 0.00198.
Also note that points with different colors may overlap,
because according to Eq.(\ref{eq-angle}), $\tan(\theta)$ is
a function of both $p_2$ (control MAF) and $p_1$ (case MAF).

{\bf Regularized $\chi$-statistic:}
Later in this section, we will establish that standardized
$\log(OR)$ is approximately equal to the square-root of
$\chi^2$ statistic, or simply $\chi$.  By following the similar definition
of regularized $t$-statistic, or SAM for significance of analysis
of microarray \citep{sam,sam-docu}, we define the regularized
$\chi$-statistic as (through standardized $\log(OR)$):
\begin{equation} 
\chi_{reg}= \frac{|\log(OR)|}{SE(\log(OR))+s_0}
= \frac{|\log n_{11}n_{22} - \log n_{12}n_{21}| }
{\sqrt{ \frac{n_{1*}}{n_{11}n_{12}}+\frac{n_{2*}}{n_{21}n_{22}}} +s_0}
=\frac{|\log p_1 q_2 -\log p_2 q_1|}
{\sqrt{ \frac{1}{p_1q_1n_{1*}}+\frac{1}{p_2q_2n_{2*}} } +s_0}
\end{equation} 
If we use $i$ to index SNPs, $\chi_{reg,i}$ contains
SNP-specific allele frequencies $(p_1, p_2)_i$, but $s_0$ is the same 
for all SNPs. The introduction of the constant $s_0$
makes the $\chi$-statistic more robust -- less sensitive to chance
fluctuation of SNP-specific standard error estimation.

Though not further used in this paper, we note that there are other ways to
define a regularized test statistic. For example, we may
use the definition of $\chi^2$-statistic and add an extra
constant in the denominator (this is parallel to a proposed
regularized $t$-test in microarray analysis \citep{baldi}):
\begin{equation}
\chi^2_{reg} = \frac{ (p_1-p_2)^2}
 {\frac{(p_1n_{1*}+p_2n_{2*})(q_1n_{1*}+q_2n_{2*})}{nn_{1*}n_{2*}}+s_0^2 }
\end{equation}
where the first term in the denominator is approximately the
variance of $p_1-p_2$; or
\begin{equation}
\chi'_{reg} = \frac{ |p_1-p_2|}
 {\sqrt{ \frac{(p_1n_{1*}+p_2n_{2*})(q_1n_{1*}+q_2n_{2*})}{nn_{1*}n_{2*}}}+s_0 }
\end{equation}
where the first term in the denominator is approximately
the SE of $p_1-p_2$.

If we select SNPs by the criterion $\chi_{reg} \ge c$, it is
equivalent to\citep{wli-review} $ y = |\log(OR)|/SE
= [ |\log(OR)|/(SE+s_0) ] \cdot [(SE+s_0)/SE] \ge c(1+s_0/\log(\theta))$.
In other words, instead of the horizontal line, the discriminant line
is a smooth curve which moves up as it is closer to the $y$-axis
(Fig.\ref{FIG1}). The regularized-$\chi$ combines information
from both $\chi^2$ test and OR.

{\bf The choice of $s_0$ in regularized $\chi$-statistic:}
The regularization constant $s_0$ in SAM for expression analysis is chosen to 
minimize the dependence of relative variation of the SAM statistic on the 
standard error \citep{sam,sam-docu,astrand}. The more 
detailed procedure in choosing $s_0$ in SAM is the following: 
genes are grouped into 100 bins by their percentile of standard errors;
within each bin, the variability of the SAM statistic is
measured by the median absolute deviation (MAD); the dependency
of relative variation of MAD on bin is calculated by
sd(MAD)/mean(MAD); the constant $s_0$ is chosen to minimize
sd(MAD)/mean(MAD). 

\begin{figure}[t]
\begin{center}
  \begin{turn}{-90}
  \epsfig{file=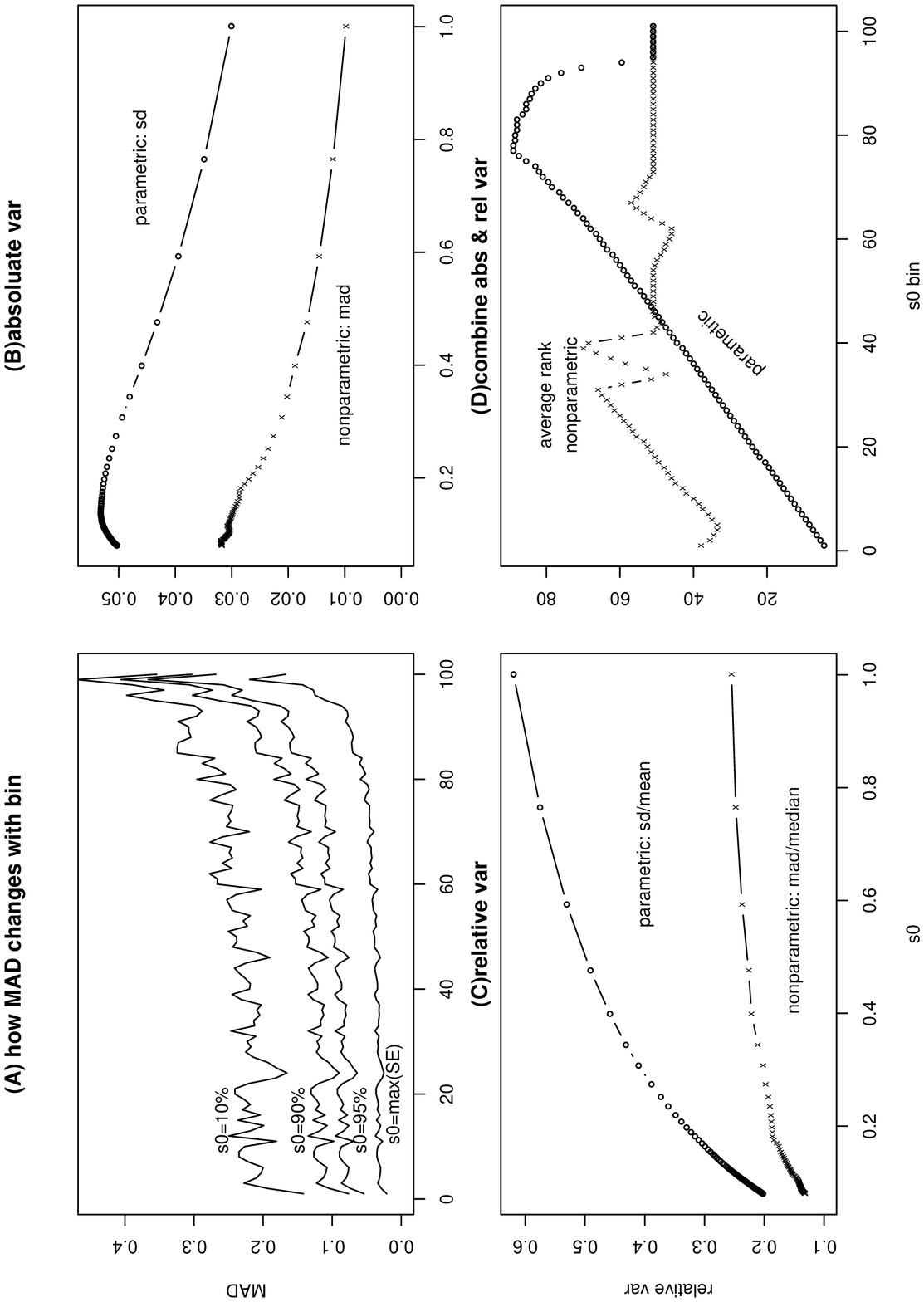, width=9cm}
  \end{turn}
\end{center}
\caption{
\label{FIG2}
}
\end{figure}

In Fig.\ref{FIG2}(A), we examine the MAD of 100 bins of SE 
values at different $s_0$'s: 10\%, 90\%, 95\%, and 100\% 
percentiles of SE. There are several observations: first,
the non-robust behavior mainly occurs at bins with large SE's.
Second, in terms of absolute variation, the choice of $s_0=\max(SE)$
seems to lead to lowest variation. Third, even if the lowest
absolute variation occurs at $s_0=\max(SE)$, because the averaged
MAD level is low, it is unclear whether the relative variation
is also low.

Fig.\ref{FIG2}(B) and (C) show indeed that the absolute
variation of MAD decreases with $s_0$, but relative
variation increases, for both the parametric and non-parametric
version of the measure of variation (sd, MAD$_{bin}$ for absolute
variation, sd/mean, MAD$_{bin}$/median for relative variation).
If the relative variation is considered, as in the original
discussion of SAM, then the $s_0=\min(SE)$ would be chosen.

Here, we consider an alternative measure of the robustness
by combining both absolute and relative variation of MAD's.
For all $s_0$ values, we rank absolute (and relative) variation
from low to high; then we add these two ranks; the $s_0$ with
the  lowest total rank is the value we use to regularize $\chi$. 
From Fig.\ref{FIG2}(D), either the min(SE) or 3\% or 4\%-percentile of
SE depending on whether the parametric or non-parametric measure
is used. For non-parametric measure, one can see that
the averaged rank is quite stable for all $s_0$ values. 

The consequence of $s_0$ is illustrated in Fig.\ref{FIG1}.
Four discriminant lines are shown for $\chi = |\log(OR)|/(SE+s_0) = \chi_0$.
The SNP filtering criterion is $\chi \ge \chi_0$. All four
lines plus the horizontal line (or $s_0=0$) selects top 70
SNPs (the corresponding $p$-value for the unregularized
$\chi^2$ test is $10^{-4}$). This can be accomplished by
tuning $\chi_0$ as the same time when various $s_0$ values
are chosen. It can be seen that with a small $s_0$ value
(0\% or 4\% percentile), there is already a great change in
the shape of discriminant line (from straight line to curve),
and many SNPs with less significant $\chi^2$ test result but
larger ORs will be selected. The discriminant line with large
$s_0$ (e.g. 100\% percentile) should probably be avoided 
because it is too different from a unregularized $\chi^2$ test.

\begin{figure}[t]
\begin{center}
  \begin{turn}{-90}
  \epsfig{file=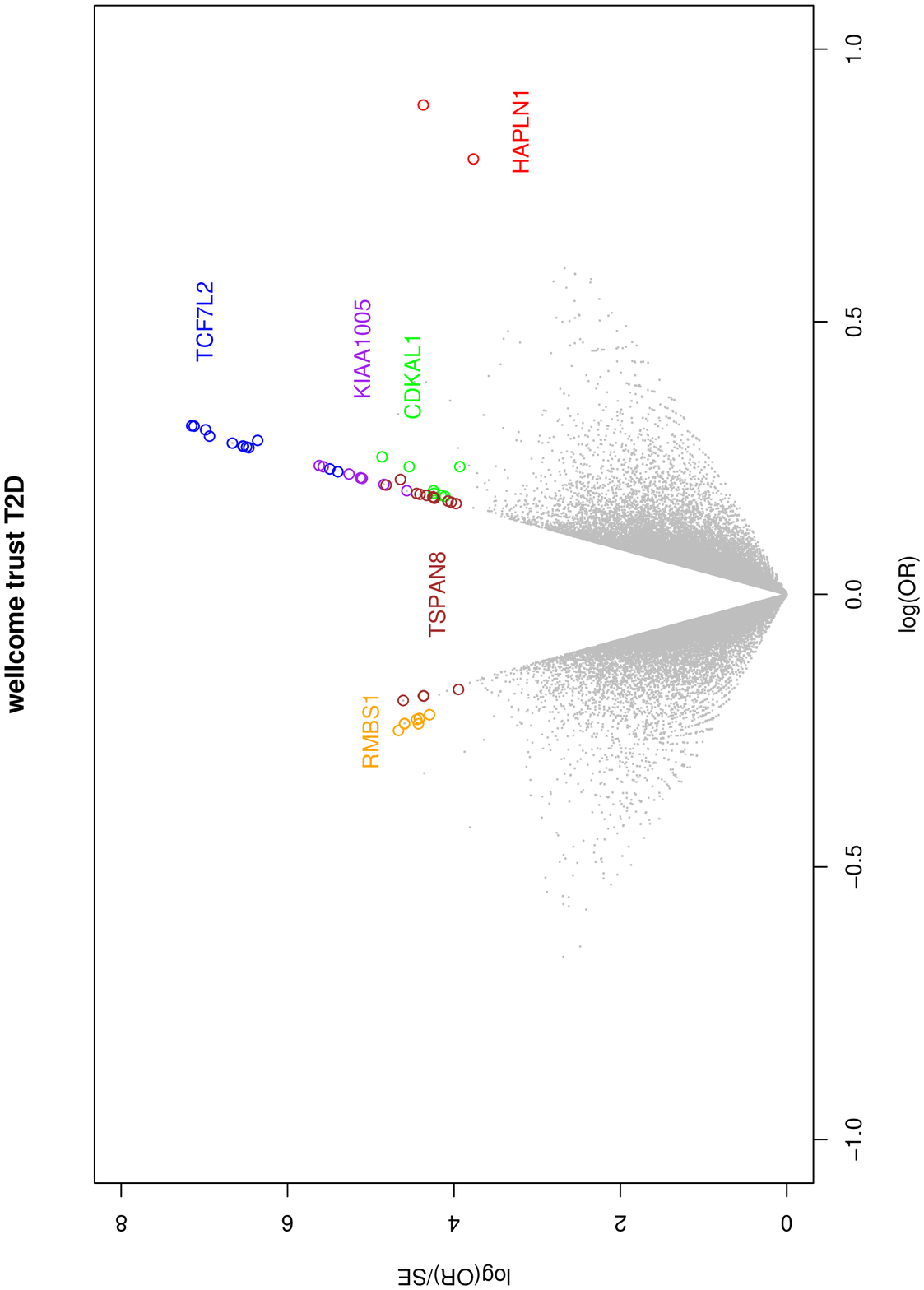, width=9cm}
  \end{turn}
\end{center}
\caption{
\label{FIG3}
}
\end{figure}

{\bf Re-examination of a published genome-wide association studies 
(GWAS) result by regularized $\chi$-statistics:} We draw
the volcano plot for 388,023 SNPs (see the Methods and Materials
section) from The Wellcome Trust Case Control Consortium type 2 
diabetes data in Fig.\ref{FIG3}. The three strongest signals
are from the gene {\sl TCF7L2} (chr10), {\sl KIAA1005} (chr16) and 
{\sl CDKAL1}
(chr6), consistent with the report in Table 3 of \citep{wellcome}.
In the subsequent validation stage, {\sl TCF7L2} and {\sl KIAA1005} signal
remains (Table 1 of \citep{t2d}) whereas {\sl KIAA1005} is dropped from
the top gene list.

Fig.\ref{FIG3} shows strikingly that the
top results in such a GWAS run are biased towards common variants,
as these genes are located at the inner envelope with the
highest possible MAFs (smallest $\theta$ angle values).
We have added two more genes further down the list:
{\sl TSPAN8} (chr12) and  {\sl RBMS1} (chr2). Interestingly, there
are SNPs on both sides of the gene {\sl TSPAN8}, and  
there are also both positive  (OR $>1$) and negative (OR $<1$)
signals. More data on {\sl TSPAN8} was reported in \citep{zeggini08},  and
the {\sl RBMS1} region has later been validated by more GWAS projects \citep{qi}

There are usually no published GWAS results for rare variants 
using the commercial genotyping arrays with the typical SNP density 
(e.g. 500k). To illustrate this in volcano plot, we highlight
the two SNPs near the gene {\sl HAPLN1} (chr5) in Fig.\ref{FIG3},
whose rankings increase the most when regularized-$\chi$
is used.  These two SNPs pass the Hardy-Weinberg equilibrium tests
in control ($p$-value=0.7) as well as in case ($p$-values=0.47, 0.52),
and they pass the differential typing test ($p$-value=0.31, 0.11),
lacking an indication of bad typing quality.  The MAF is increased from 
0.0065 in control to 0.015 in case, with $\chi^2$-statistic of 
20.36, 14.93 ($p$-values= 6$\times 10^{-6}$), 1$\times 10^{-4}$, 
and ORs 2.5, 2.2. However, these two SNPs would not have passed 
the filtering in the original WTCCC analysis because the MAF 
is lower than 0.01. Using volcano plot and regularized-$\chi$, 
these rare variants are easily highlighted and deserve further attention.

{\bf Other potential choices of $x$- and $y$-axis of volcano plots:} 
Besides the log-odds-ratio, other candidate for
the unstandardized variables for the $x$-axis include minor allele
frequency difference $d_{MAF}$ and the fixation index $F_{st}$.
The MAF difference $d_{MAF}= (n_{11}n_{22}- n_{12}n_{21})/(n_{1*}n_{2*})$
may look very different from $\log(OR)$, but under the null hypothesis (i.e.
zero allele frequency difference), the two measures are related, because: 
\begin{equation}
\label{eq-delta}
\log(OR) = \log \left( 1+\frac{d_{MAF}}{p_2} \right) \left( 1+\frac{d_{MAF}}{q_1} \right)
\approx \frac{d_{MAF}}{p_2} + \frac{d_{MAF}}{q_1} + O(d_{MAF}^2)
= \frac{d_{MAF}}{(1-p_1)p_2} + O(d_{MAF}^2)
\end{equation}

\begin{figure}[t]
\begin{center}
  \begin{turn}{-90}
  \epsfig{file=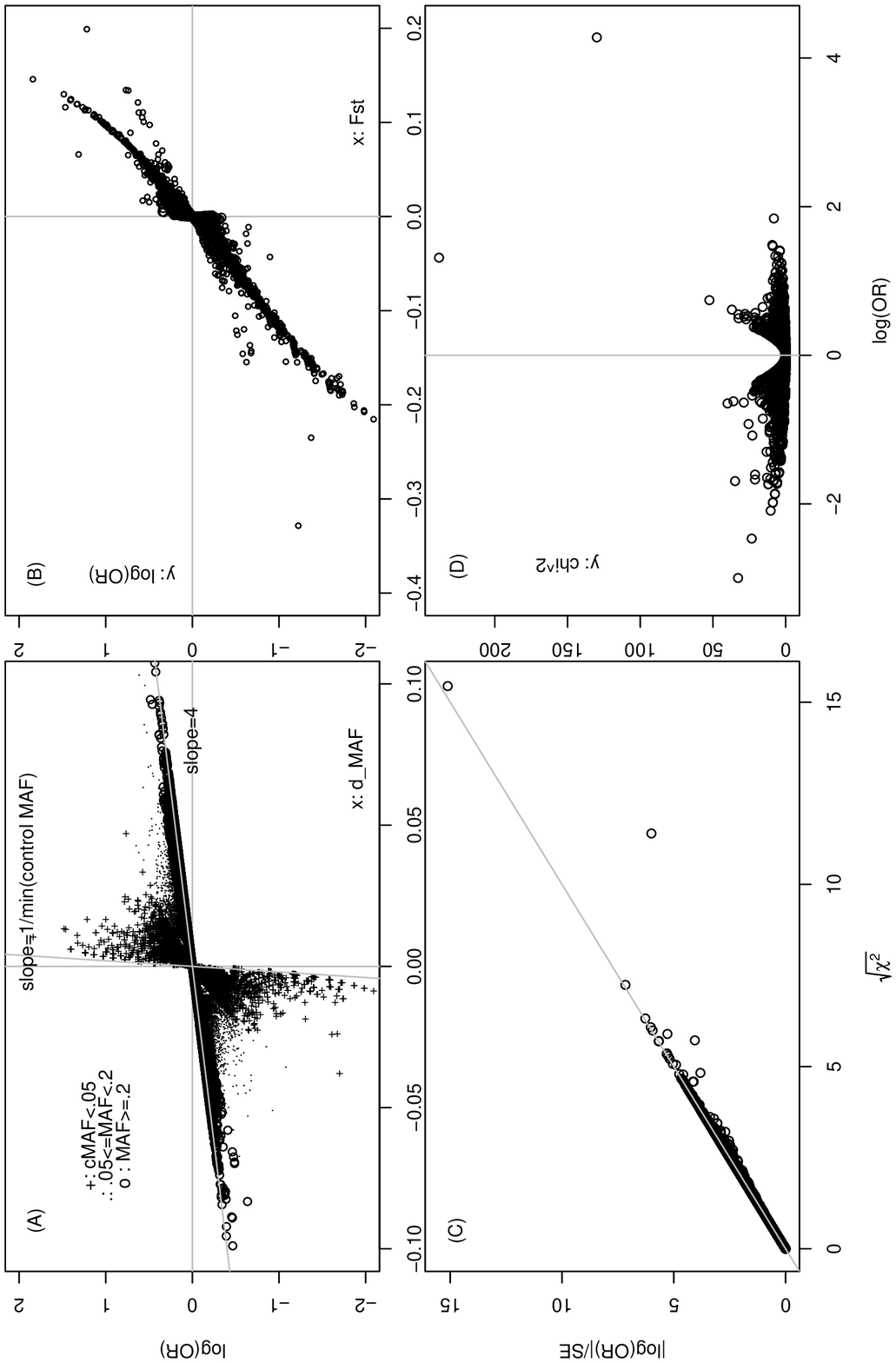, width=9cm}
  \end{turn}
\end{center}
\caption{
\label{FIG4}
}
\end{figure}

Thus, if $d_{MAF}$ is far from zero, there is no simple
relationship between the two. Fig.\ref{FIG4}(A) shows the
existence of two distinct branches in the $d_{MAF}$ vs. $\log(OR)$
scatter plot. The first branch is for rare-allele SNPs (low $p_1,p_2$ value),
which more or less trace the line $\log(OR) \sim p_2^{-1} d_{MAF}$.
The second branch is for common alleles ($p_1 \approx p_2 \approx 0.5$), where
$\log(OR) \sim 4 d_{MAF}$. Both approximations can be obtained
from Eq.(\ref{eq-delta}).  
SNPs with low MAF are more likely to achieve high OR values, 
but never high $d_{MAF}$; whereas SNPs with common MAF 
tend to have large $d_{MAF}$, but only limited OR.
Note that SNPs which rank high by $\chi^2$ test result shown
in Fig.\ref{FIG3} belong to the common variant branch, whereas
those ranking relatively high in regularized-$\chi$ (Fig.\ref{FIG3})
tend to belong to the rare variant branch.

The fixation index $F_{st}$ is highly correlated with
$\log(OR)$ (Fig.\ref{FIG4}(B)).  Interestingly, points (SNPs) with different
MAF values overlap with each other on Fig.\ref{FIG4}(B), thus
not stratified by MAF (result not shown). It can be shown that
\begin{equation}
F_{st}= \frac{d_{MAF}^2}{(p_1+p_2)(2-p_1-p_2)},
\end{equation}
so $F_{st}$ scales as the square of allele frequency differences.

Here we used the standardized $\log(OR)$ as the $y$-axis, but could instead also
have used the $\chi^2$-statistic.  In fact, the two
are very similar (see Fig.\ref{FIG4}(C)), and both are expected 
to approximate a standard normal distribution.
Fig.\ref{FIG4}(D) and Fig.\ref{FIG1} show two versions
of a volcano plot, with the former uses $\chi^2$
vs. $\log(OR)$, and second uses $\log(OR)/SE(\log(OR))$ 
vs. $\log(OR)$. The difference between the two is mostly 
due to the fact that $\chi^2$ is the square of a normally 
distributed variable, so that straight lines in 
Fig.\ref{FIG1} become parabola in Fig.\ref{FIG4}(D).

\section*{Discussion}
\indent

Like any graphical representation of data or
analysis results, such as effect size vs. sample size in the funnel 
plot \citep{egger}, true positive rate vs. false positive rate 
in receiver operating characteristic (ROC) curve \citep{roc},  etc.,
the introduction of volcano plot to the genetic association
studies brings in new perspectives. The role of MAF in balancing
test $p$-value and OR, and in the biased selection of
variants in GWAS, can be easily concluded from the volcano plot. 

The idea of regularized-$\chi$ is the same as that
of regularized-$t$ (or SAM) in microarray analysis:
the avoidance of over-confidence in the ability to exactly
estimate variances. The
consequence is that those SNPs (or genes in microarray data)
with extremely good test result (due to low standard
error estimations) -- but mediocre signal strength --
move down in the ranking list.

The goal of the current paper is to introduce
the concept of regularized-$\chi$, whereas more details have to be worked out
in future publications. For example,
the choice of $s_0$ here is not based on a
solid theoretical ground. However, the same comment may also be made
on the SAM in microarray analysis. And we have shown that for $s_0$
being non-zero is more important than having a specific value.
Also, we mainly focus on the effect of regularization on the ranking
order of SNPs, thus the choice of the threshold value and the resulting
distribution of type I and type II error rates has not been discussed.

Regularized-$\chi$ can be applied to published GWAS results
even when only the summary statistics are available.  We
have known in Fig.\ref{FIG4}(C) that the square-root of
$\chi^2$-statistic is approximately equal to the standardized
$\log(OR)$, or $\log(OR)/SE(\log(OR))$. Consequently,
SE(log(OR)) is equal to $\log(OR)/\sqrt{\chi^2}$.
Even if we may not know the distribution of SE(log(OR)) 
for all SNPs when a publication only provides the 
top-ranking results, these SNPs tend to have lower value
of SE(log(OR)); the minimum of them could be the 
$s_0$ value in defining regularized-$\chi$.

Unlike the regularized-$t$ in microarray analysis, in genetic
association analysis we have a clear understanding of the cause for a low 
standard error. This can be seen from Fig.\ref{FIG1} where the points/SNPs 
forming small angles with the $y$-axis, thus having low standard errors, 
are all common variants with higher MAFs. Indeed, common SNPs have more 
statistical power than rare variants, but the true disease susceptibility
genes with low allele frequencies are likely to be missed if $p$-values
are used as the filtering criterion.  The purpose or consequence of 
regularized-$\chi$ then becomes clear: it puts the signal originating 
from rare variants as measured by OR in the context of common variant
association signals.

On the practical side, this effect of regularized-$\chi$ to select rare 
variants can come in conflict with the quality control, because
genotyping errors can be mistaken as rare variants. 
Points/SNPs with the lowest MAFs form the bottom layer
of the envelope in Fig.\ref{FIG1}, and the only way these would
pass the regularized-$\chi$ threshold is to have large OR values. 
In fact, the OR could be infinity when one of the allele count is zero
(though in principle, it could be avoided by a Yate's correction). 
As a result, requiring a minimum number of minor allele 
(in both case and control group) to be included in the dataset 
can be an effective way to exclude low-quality SNPs to be selected. 
However, as sample size increases and genotyping technology matures, this
becomes less of a concern.  Ultimately, appropriate filtering threshold for 
MAF depends on the genotyping technology (e.g. microarray versus
exome sequencing) and its error rate.

It is well known that genetic association signals from rare variants
using array-based genotyping data is difficult. With the 
low density (500k) SNPs and low number of samples, rare 
disease-gene-containing haplotype may not be tagged effectively. 
However, with the next-generation
sequencing (NGS) data, rare variances are called with
more confidence, and we expect the volcano plot could play
an important role in the analysis of such data.

\section*{Acknowledgments}

W.L. and J.F. acknowledges the support from the
Robert S. Boas Center for Genomics and Human Genetics.

\newpage

FIGURE CAPTIONS

\normalsize

Fig.1: Volcano plot of 38735 SNPs located in chromosome 6 for a GWAS 
for an autoimmune disease with 809 cases and 505 controls. 
The angle $\theta$ is related to the
standard error of $\log(OR)$ by the equation: $\tan(\theta)=SE(\log(OR))$.
The colors red, orange, purple,and  blue label SNPs with control
MAFs in the intervals of (0.00198, 0.01), (0.01, 0.05), (0.05,0.2), and (0.2, 0.5).
The horizontal line corresponds to $\sqrt{\chi^2}=3.89$ or
$p$-value equal to $10^{-4}$. The threshold
$\chi \ge 3.89$ filters 70 SNPs. The threshold for regularized $\chi$ with
$s_0=0.08$ (minimum of SE), $s_0=0.08049$ (4\% percentile), $s_0=0.2194957$
(90\% percentile), and $s_0=1.00086$ (maximum of SE) are also shown,
where the $\chi_0$ ($\chi \ge \chi_0$) is chosen so that exactly
70 SNPs are filtered.

Fig.2: 
(A) MAD (median of absolute deviation) of regularized $\chi$'s
in 100 bins of SE(log(OR))'s at 4 $s_0$ values: $s_0=0.081$ (10\% percentile),
$s_0=0.2194957$ (90\% percentile), $s_0=0.344$ (95\% percentile),
and $s_0=1.00086$ (maximum or 100\% percentile).
(B) Two measures of absolute variation of MAD's in (A) along
bins: standard deviation (sd(MAD)) and median of absolute deviation
(MAD$_{bin}$(MAD) multiplied by 1.4826), as a function of $s_0$.
(C) Two measures of relative variation of MAD's in (A) along
bins: coefficient of variation (sd(MAD)/mean(MAD)) and MAD$_{bin}$(MAD)/median(MAD),
as a function of $s_0$.
(D) Sum of rank of absolute variation in (B) and rank of
relative variation in (C) divided by 2. The ranking is from low
to high values. The $x$-axis is the bin number for $s_0$'s.
  
Fig.3: The volcano plot for The Wellcome Trust Case Control
Consortium (WTCCC)'s type 2 diabetes (T2D) data, with 1924
cases and 2938 controls. Only a small portion of the 388,023 SNPs
are shown as the background, with those on the following
genes are highlighted: {\sl TCF7L2} (chr10, blue), {\sl KIAA1005} (chr16, purple),
{\sl CDKAL1} (chr6, green), {\sl RBMS1} (chr2, orange, on the negative branch),
{\sl TSPAN8} (chr12, brown, on both positive and negative branch),
and  {\sl HAPLN1} (chr5, red, rare variant).

Fig.4: 
(A) Scatter plot of $d_{MAF}$ ($x$-axis) and $\log(OR)$ ($y$-axis).
Points far away from the origin are not plotted.
Points (SNPs) are stratified by MAF in control group:
crosses for low MAF (MAF $< 0.05$), circles for high MAF
(MAF $> 0.2$), with all other points represented by dots.
The two straight lines seem to envelope all points:
one with slope 4 which traces common-allele SNPs,
and another with slope $1/\min(p_2)$ which
traces rare-allele SNPs.
(B) Scatter plot of $F_{st}$ ($x$-axis) and $\log(OR)$ ($y$-axis).
(C) scatter plot of square-root of $\chi^2$-statistics ($x$)
and standardized $\log(SE)$ in absolute value, 
$|\log(OR)|/SE(\log(OR))$ ($y$).
(D) volcano plot with $\log(OR)$ as $x$, $\chi^2$-statistics
as $y$.

\newpage

\end{document}